\documentclass[aps,preprint,showpacs]{revtex4}

\begin{document}

\title{Dynamics and transformations of Josephson vortex lattice in
layered superconductors}

\author{D.A.Ryndyk, V.I.Pozdnjakova, I.A.Shereshevskii, N.K.Vdovicheva}

\affiliation{Institute for Physics of Microstructures,
Russian Academy of Sciences \\
GSP-105, Nizhny Novgorod 603600, Russia}

\email{ryn@ipm.sci-nnov.ru}

\date{25 of April, to be published in Phys. Rev. B}

\begin{abstract}
We consider dynamics of Josephson vortex lattice in layered
superconductors with magnetic, charge (electrostatic) and
charge-imbalance (quasiparticle) interactions between interlayer
Josephson junctions taken into account. The macroscopic dynamical
equations for interlayer Josephson phase differences, intralayer
charge and electron-hole imbalance are obtained and used for
numerical simulations. Different transformations of the vortex
lattice structure are observed. It is shown that the additional
dissipation due to the charge imbalance relaxation leads to the
stability of triangular lattice.
\end{abstract}


\maketitle

The investigation of the resistive state of layered high-T$_c$
superconductors in the external magnetic field parallel to the layers
is the important task of modern experiment and theory. The c-axis
resistivity in this case is determined by Josephson flux flow. Thus,
the nonstationary properties of HTSC can be investigated in wide
parameter range of temperatures, fields and currents. In the recent
experiments \cite{Lee95,Lee97,Irie98,Krasnov99} flux flow states in
low magnetic fields are observed and in
Ref.\,\cite{Hechtfischer97a,Hechtfischer97b} -- in high fields up to
3.5 T. The number of peculiarities is discovered. In particular, in
Ref.\,\cite{Lee97,Krasnov99} multiple flux flow branches, associated
with different vortex modes, are observed. In
Ref.\,\cite{Irie98,Krasnov99} it is shown that two universal flux
flow regimes with different $V/H$ take place at different magnetic
fields and transformation from the first state to the second is
possible with current rise at intermediate fields. Finally, in
Ref.\,\cite{Hechtfischer97b} broad-band microwave emission is
measured with frequencies much less than Josephson frequency
$\omega_J=2eV/N\hbar$. All this phenomena are related to the
dynamical transformations of Josephson vortex lattice (JVL) and
vortex interaction with linear plasma modes (see also Ref.\,
\cite{Sakai93,Bulaevskii94,Kleiner94a,Kleiner94b,Ustinov96,Artemenko97b,Ustinov98a,
Volkov98,Machida00,Koshelev00,Kleiner00}).

On the other hand, there is the long-standing problem to obtain the
coherent electromagnetic emission from stacked Josephson junction.
The most simple way to do this is to achieve ''in-phase'' (square)
arrangement of vortices. Although at zero external current the
triangular JVL is favorable, it was shown for two-junction stack
theoretically and experimentally \cite{Ustinov96,Volkov98} and for
multi-junction stack theoretically
\cite{Kleiner94a,Kleiner94b,Ustinov98a,Machida00,Koshelev00} that
moving square JVL can be stable. However no indications of this
in-phase regime were found in experiments with HTSC. The reason for
this discrepancy may be that only magnetic coupling between layers
have been taken into account in calculations.

In
\cite{Bulaevskii96a,Bulaevskii96b,Koyama96,Artemenko97,Ryndyk98,Ryndyk99}
it was also suggested that in the case of thin superconducting layers
some nonequilibrium mechanisms are to be taken into account, namely
electrical charging of layers (charge effect) and nonequilibrium
quasiparticles (charge-imbalance effect). To investigate this effects
we obtain macroscopic dynamical equations for interlayer Josephson
phase differences with magnetic, charge and charge-imbalance
interactions taken into account.

Following \cite{Sakai93,Bulaevskii94} we use London expression for
the longitudinal in-layer supercurrent neglecting longitudinal
in-layer electric field ${\bf E}_n^{\parallel}$ and normal current
(with accuracy \mbox{$(\sigma_{ab}/\sigma_c)\gamma^{-2}\ll\!1$},
$\gamma=\lambda_c/\lambda_{ab}$ is the anisotropy parameter)
\begin{equation}
{\bf j}_n^{\parallel}({\bf r})=-\frac{c\Phi_0}{8\pi^2\lambda^2}\left[
{\bf\nabla}_{\parallel}\theta_n({\bf r})+\frac{2\pi}{\Phi_0}{\bf
A}_n({\bf r})\right],
\end{equation}
together with the equation ${\bf B}={\rm rot}{\bf A}$ and the
definition of invariant Josephson phase
$\varphi_{n,n+1}=\theta_{n+1}-\theta_n-\frac{2\pi}{\Phi_0}
\int_{nd}^{(n+1)d} A_zdz.$ Integrating {\bf A} over closed contour
one obtains
\begin{equation}
{\bf B}\times{\bf z}_0=\frac{\Phi_0}{2\pi d}{\bf\nabla}_{\parallel}
\varphi_{n,n+1}-\frac{4\pi\lambda^2}{cd}({\bf j}_{n+1}^{\parallel}
-{\bf j}_n^{\parallel}),
\end{equation}
here ${\bf B}$ is the averaged interlayer magnetic field, ${\bf z}_0$
is the unity vector perpendicular to the layer and $d$ is the
interlayer distance, layers assumed to be thin ($d\ll\lambda_L$).

Then, using Maxwell equation ${\rm rot}{\bf B}
=\frac{\epsilon}{c}\frac{\partial\bf E}{\partial t}+
\frac{4\pi}{c}{\bf j},$ projected on to z-axes, we obtain
\begin{equation}
\frac{\epsilon}{c}\frac{\partial E_{n,n+1}}{\partial t}+
\frac{4\pi}{c}j_{n,n+1}=\frac{\Phi_0}{2\pi
d}{\bf\nabla}^2_{\parallel}
\varphi_{n,n+1}-\frac{4\pi\lambda^2}{cd}{\bf\nabla}_{\parallel} ({\bf
j}^{\parallel}_{n+1}-{\bf j}^{\parallel}_n).
\end{equation}
Taking use of the continuity equation $\frac{\partial\rho_n}{\partial
t}+{\bf\nabla}_{\parallel} {\bf
j}^{\parallel}_n+\frac{j_{n,n+1}-j_{n-1,n}}{d_0}=0$ and Maxwell
equation $\epsilon{\rm div}{\bf E}=4\pi\rho$ in integral form
$\rho_n=\frac{\epsilon}{4\pi d_0}(E_{n,n+1}-E_{n-1,n})$ one obtains
finally the equation
\begin{equation}
\frac{c\Phi_0 d_0}{16\pi^2\lambda^2}{\bf\nabla}^2_{\parallel}
\varphi_{n,n+1}=\left(1+\frac{dd_0}{2\lambda^2}\right)
j^*_{n,n+1}-0.5(j^*_{n-1,n}+j^*_{n+1,n+2}),
\end{equation}
\begin{equation}
j^*_{n,n+1}=j_{n,n+1}+\frac{\epsilon}{4\pi}\frac{\partial E_{n,n+1}}
{\partial t}-j_{ext} =
j_{n,n+1}+\frac{\epsilon}{4\pi d}\frac{\partial
V_{n,n+1}} {\partial t}-j_{ext},
\end{equation}
where $j_{ext}$ is the external current in overlapping geometry (for
the details of introducing the external current in different
geometries see Ref.\,\cite{Sakai93}).

To find the interlayer Josephson current $j_{n,n+1}$ we use the
theory of nonequilibrium Josephson effect developed recently
(\cite{Bulaevskii96a,Bulaevskii96b,Koyama96,Artemenko97,Ryndyk98,Ryndyk99}
and references therein).  We take into account that in nonequilibrium
state there is nonzero invariant potential
\begin{equation}
\label{phi}
\Phi_n(t)=\phi_n+(\hbar/2e)(\partial\theta_n/\partial t),
\end{equation}
where $\phi_n$ is the electrostatic potential and $\theta_n$ is the
phase of superconducting condensate, $\Phi=0$ in the equilibrium
state.

An ordinary Josephson relation $(d\varphi/dt)=(2e/\hbar)V$ between
the Josephson phase difference $\varphi_{n,n+1}$ and voltage
$V_{n,n+1}=E_{n,n+1}d$ is violated. Instead, we have (from
(\ref{phi}))
\begin{equation}
\label{NJR}
\frac{\partial\varphi_{n,n+1}}{\partial t}=
\frac{2e}{\hbar}V_{n,n+1}+\frac{2e}{\hbar} (\Phi_{n+1}-\Phi_n).
\end{equation}

The charge density inside a superconducting layer is
\begin{equation}
\label{rho}
\rho_n=-2e^2N_0(\Phi_n-\Psi_n),
\end{equation}
where the first term is the charge of superconducting electrons
determined by the shift of the condensate chemical potential
$\delta\mu_s=-e\Phi$ and the second term is the quasiparticle charge,
described by potential $\Psi_n$ (details can be found in
Ref.\,\cite{Ryndyk99}). Finally, the equation for the charge
imbalance can be written in the form corresponding to the generalized
nonstationary Ginzburg-Landau theory \cite{Ryndyk99}
\begin{equation}
\label{eq_psi2}
\frac{\partial\Psi_i}{\partial t}=
(1-\Gamma)\frac{\partial\Phi_i}{\partial
t}+2\nu_t\frac{\hbar}{2e}\left(\frac{\partial\varphi_{i-1,i}}{\partial t}
-\frac{\partial\varphi_{i,i+1}}{\partial t}\right)
+2\nu_t(\Psi_{i-1}+\Psi_{i+1}-2\Psi_i)-\tau_q^{-1}\Psi_i,
\end{equation}
where $\eta=2\nu_t\tau_q$ is the parameter of disequilibrium,
\mbox{$\tau_q$} is the well-known charge-imbalance relaxation time,
$\nu_t=(4e^2N_0RSd_0)^{-1}$ is the ''tunnel frequency'', $R$ is the
normal resistivity of the tunnel junction, $V=Sd_0$ is the volume of
the superconducting layer, $N_0=mp_F/2\pi^2\hbar^3$ is the density of
state (or, instead, $N_f=N_0d_0$ can be considered as 2D density of
states), $\Gamma\ll 1$ is the parameter dependent on the details of
microscopic theory.

In the same approximation we derive the expression for the nonequilibrium
interlayer current \cite{Ryndyk99}
\begin{equation}
\label{jeq}
j_{i,i+1}=j_c \sin\varphi_{i,i+1} +
\frac{\hbar}{2eR}\frac{\partial\varphi_{i,i+1}}{\partial t}+
\frac{\Psi_{i}-\Psi_{i+1}}{R}.
\end{equation}

Finally, in dimensionless form we obtain
\begin{equation}
\label{3D1}
{\bf\nabla}^2_{\parallel}\varphi_{i,i+1}=
j^*_{i,i+1}-s(j^*_{i-1,i}+j^*_{i+1,i+2}),
\end{equation}
\begin{equation}
j^*_{i,i+1}=\beta\frac{\partial^2\varphi_{i,i+1}}{\partial\tau^2}+
\frac{\partial\varphi_{i,i+1}}{\partial\tau}+
\sin\varphi_{i,i+1}+\psi_i-\psi_{i+1}
+\beta\left(\frac{\partial\mu_i}{\partial\tau}-
\frac{\partial\mu_{i+1}}{\partial\tau}\right)-j_{ext},
\end{equation}
\begin{equation}
\alpha'\frac{\partial\psi_i}{\partial\tau}
+\psi_i+\eta(2\psi_i-\psi_{i-1}-\psi_{i+1})
=\eta\left(\frac{\partial\varphi_{i-1,i}}{\partial\tau}
-\frac{\partial\varphi_{i,i+1}}{\partial\tau}\right)
+\alpha'(1-\Gamma)\frac{\partial\mu_i}{\partial\tau},
\end{equation}
\begin{equation}
\label{3D4}
\mu_i+\zeta(2\mu_i-\mu_{i-1}-\mu_{i+1})
=\psi_i+\zeta\left(\frac{\partial\varphi_{i-1,i}}{\partial\tau}
-\frac{\partial\varphi_{i,i+1}}{\partial\tau}\right).
\end{equation}
where  $\mu(t)=\Phi(t)/V_c$, $\psi(t)=\Psi(t)/V_c$,
$V_c=\hbar\omega_c/2e$, ¨ $\alpha'=\tau_q\omega_c$,
$\beta=(\hbar\epsilon\omega_c^2S)/(8\pi edj_c)$, $\lambda_J^2=(\hbar
c^2d_0)/(16\pi ej_c(\lambda^2+dd_0))$, $\zeta=(\epsilon
r_d^2)/(d_0d)$, $\omega_c=2eRj_c/\hbar$, $\tau=\omega_c t$,
$\tilde{x}=x/\lambda_J$, $s=0.5/(1+(dd_0)/(2\lambda^2)))$.

This system of equations is a generalization of well-known magnetic
coupling model of layered superconductors
\cite{Sakai93,Bulaevskii94}. Neglecting electron-hole imbalance
($\eta=0$, $\Psi_n=0$), we derive the system of equations obtained in
Ref. \cite{Machida00}, and neglecting the charge effect ($\zeta=0$,
$\Psi_n=\Phi_n$) we derive the system of equations similar to that of
\cite{Bulaevskii96a,Bulaevskii96b,Melnikov96}.

Before further study, the Eqs. (\ref{3D1})-(\ref{3D4}) must be added
with boundary conditions, which we write as
\begin{equation}
\label{3DBC}
\begin{array}{c}
j^*_{-1,0}=j^*_{N,N+1}=0,\\
\left.
\displaystyle{\frac{\partial\varphi_{i,i+1}}{\partial x}}\right|_{0,L}=B,\\
\psi_{0}=\psi_{N}=0,\\ \left.
\displaystyle{\frac{\partial\psi_{i,i+1}}
{\partial x}}\right|_{0,L}=0,\\
\mu_{0}=\mu_{N}=0.
\end{array}
\end{equation}

For numerical solution of the equations (\ref{3D1})-(\ref{3D4}) with
boundary conditions (\ref{3DBC}) we introduce the new functions
$\hat{\varphi},\ v,\ z$, defined by the relations

\begin{equation}
\label{Var}\begin{array}{c}
                \hat{\varphi}_i(x,t)=\varphi_{i,i+1}(x,t)-Bx,\\
        v_i(x,t)=
    \displaystyle{\frac{\partial\hat{\varphi}_i(x,t)}{\partial t}}+
        \psi_i(x,t)-\psi_{i+1}(x,t),\\
        z_i(x,t)=
        \varepsilon
    \displaystyle{\frac{\partial\hat{\varphi}_i(x,t)}{\partial t}}+
        \alpha'(\psi_i(x,t)-\psi_{i+1}(x,t)),\\
        i=0,...N-1,\ \varepsilon=\alpha'(1-\Gamma).
       \end{array}
\end{equation}
Let us denote now by $U$ the column vector-function of kind
\begin{equation}
\label{UVar}
    U=\left(\begin{array}{c}
        \hat{\varphi}\\
        v\\
        z\\
       \end{array}\right),
\end{equation}
by $\hat{A}$ the operator matrix of kind
\begin{equation}
\label{AOp}
    \hat{A}=\left(\begin{array}{ccc}
                0 & \frac{\alpha'\beta}{\alpha'-\varepsilon} & -\frac{\beta}{\alpha'-\varepsilon} \\
                0&-1+\zeta\Delta_d&0\\
        0&-\varepsilon(1-\frac{\beta}{\alpha'-\varepsilon})-
                   \frac{\beta\gamma\varepsilon}
        {\alpha'-\varepsilon}\nabla_\parallel^2
               +\beta\eta\Delta_d
         &\frac{\beta\gamma}{\alpha'-\varepsilon}\nabla_\parallel^2
             -\frac{\beta}{\alpha'-\varepsilon}\\
       \end{array}\right),
\end{equation}
(here $\Delta_d$ is the difference Laplasian, $(\Delta_d
f)_i=f_{i-1}-2f_i+f_{i+1}$) and by $F$ the column vector of kind

\begin{equation}
\label{Fvec}
    F=\left(\begin{array}{c}
        0\\
                (1-\zeta\Delta_d)(j^*+j_{ext}-\sin(\varphi))\\
                \varepsilon(j^*+j_{ext}-\sin(\varphi))\\
       \end{array}\right).
\end{equation}

Then we can rewrite the above system of equations in the following
"operator" form:
\begin{equation}
\label{OpEc}\begin{array}{c}
    \beta
    \displaystyle{\frac{\partial U}{\partial t}}=\hat{A}U+F,\\
        (s\Delta_d+2s-1)j^*=-\nabla_\parallel^2\hat{\varphi},\\
    \end{array}
\end{equation}
with appropriate homogeneous boundary conditions. For numerical
solution of the equation (\ref{OpEc}) we use, after sampling in
$x$-direction, the semi-implicit scheme of form
\begin{equation}
\label{Sch}
    U(t+\Delta t)=
    2(E-\frac{\Delta t}{2\beta}\hat{A})^{-1}(U(t)+
    \frac{\Delta t}{\beta}F(t))-U(t)-\frac{\Delta t}{\beta}F(t),
\end{equation}
in which all  needed inverse operators can be easily evaluated by
standard sweep method. The transformation from vector $U$ to initial
variables is evident.

The developed program gives us the possibility to observe the state
of system during calculation process in convenient graphical form. We
can therefore find many different variants of system dynamics and
vortex lattice structures.

The most part of simulations has been carried out for the case of
high magnetic fields $H>H^*=\Phi_0/\gamma t^2$, where
$\Phi_0=\pi\hbar c/e$ is the flux quantum,
$\gamma=\lambda_c/\lambda_{ab}$ is the anisotropy parameter, and
$t=d+d_0$ is the period of the structure. At this fields triangular
vortex lattice is formed in the static case.

The results of numerical simulations can be shortly described in the
following way. At small currents vortex lattice remains triangular.
At high currents the situation depends on dissipation. If interlayer
dissipation is strong enough ($\beta<1$) then triangular lattice is
stable at all currents. Otherwise, we observed various
transformations of lattice structure (further $\beta=100$ and
$s=0.48$ are considered).

(i) At pure magnetic coupling ($\zeta=0$, $\eta=0$) we obtain the
typical picture of JVL transformations. The example is shown on
Fig.\,1. We consecutively change current from zero to some high value
($a$-$f$) and then decrease it up to zero again ($g$-$l$). The square
lattice is clear seen on Fig.\,$d$,$e$,$i$, and $j$. Low current
regimes $a$,$l$, as well as high current one $f$,$g$ are triangular
(or close to it). Regimes $b$,$c$ and $k$ are "inhomogeneous", --
velocities of vortex chains are different in different layers. These
regimes are also "breathing" due to strongly excited nonlinear plasma
waves. We find that such breathing modes can be triangular or square
in average. Note that different regimes exist at the same current
that can lead to hysteresis on VCC.

(ii) If we take charge coupling into account (Fig.\,2, $\zeta=1$,
$\eta=0$), the picture of JVL transformations is qualitatively the same
but the regimes with square lattice are shifted to higher currents.

(iii) For charge-imbalance (quasiparticle) coupling the result is
qualitatively different. We have found that if the time of
charge-imbalance relaxation is large enough (approximately
$\alpha'=\tau_q\omega_c>1$) than the triangular lattice become stable
even if the parameter of disequilibrium $\eta$ is small. The example
is shown on Fig.\,3 ($\zeta=0.1$, $\eta=0.1$, $\alpha'=100$). There
is no JVL transformations at this parameters although both charge and
charge-imbalance couplings are weak. The origin of this effect is the
additional dissipation due to the charge-imbalance relaxation. The
detailed theory will be considered in other publication.

In conclusion, we derive macroscopic dynamical equations for layered
superconductors with magnetic, charge and nonequilibrium
quasiparticle interactions taken into account. Many dynamical regimes
are observed, in particular triangular, square, inhomogeneous and
breathing modes. We established that additional dissipation due to
the charge-imbalance relaxation can prevent JVL transformations and
make triangular lattice stable at all currents.

This work is supported by the Russian Foundation for Basic Research,
Grants No. 00-02-16528, 99-02-16188 and 00-15-96734 ("Leading
Scientific Schools"). The authors thanks Prof. A. Andronov, Prof. J.
Keller, Prof. R. Kleiner, Dr. V. Kurin, Dr. A. Mel'nikov,  and Dr.
Ch. Preis for valuable discussions.

\newpage
\section*{Figure captions}

Fig.1. The structure of Josephson vortex lattice ($\sin\varphi_n$ are
shown) with $\beta=100$, $s=0.48$, $\eta=0$, $\zeta=0$,
$j=0.2,0.4,0.6,0.8,1,1.2$. Current is increased ($a$-$f$) and
then decreased ($g$-$l$).

Fig.2. The structure of Josephson vortex lattice with $\beta=100$,
$s=0.48$, $\eta=0$, $\zeta=1$.

Fig.3. The structure of Josephson vortex lattice with $\beta=100$,
$s=0.48$, $\eta=0.1$, $\zeta=0.1$, $\alpha'=100$, $\Gamma=0.01$.


\begin{thebibliography}{10}

\bibitem{Lee95}
J. Lee, J. Nordman, and G. Hohenwarter, Appl. Phys. Lett. {\bf 67},  1471
  (1995).

\bibitem{Lee97}
J. Lee {\it et~al.}, Appl. Phys. Lett. {\bf 71},  1412  (1997).

\bibitem{Irie98}
A. Irie, Y. Hirai, and G. Oya, Appl. Phys. Lett. {\bf 72},  2159  (1998).

\bibitem{Krasnov99}
V. Krasnov, N. Mros, A. Yurgens, and D. Winkler, Phys. Rev. B {\bf 59},  8463
  (1999).

\bibitem{Hechtfischer97a}
G. Hechtfischer {\it et~al.}, Phys. Rev. B {\bf 55},  14638  (1997).

\bibitem{Hechtfischer97b}
G. Hechtfischer, R. Kleiner, A. Ustinov, and P. Muller, Phys. Rev. Lett. {\bf
  79},  1365  (1997).

\bibitem{Sakai93}
S. Sakai, P. Bodin, and N. Pedersen, J. Appl. Phys. {\bf 73},  2411  (1993).

\bibitem{Bulaevskii94}
L. Bulaevskii {\it et~al.}, Phys. Rev. B {\bf 50},  12831  (1994).

\bibitem{Kleiner94a}
R. Kleiner {\it et~al.}, Phys. Rev. B {\bf 50},  3942  (1994).

\bibitem{Kleiner94b}
R. Kleiner, Phys. Rev. B {\bf 50},  6919  (1994).

\bibitem{Ustinov96}
A. Ustinov and H. Kohlstedt, Phys. Rev. B {\bf 54},  6111  (1996).

\bibitem{Artemenko97b}
S. Artemenko and S. Remizov, JETP Letters {\bf 66},  853  (1997), [Pis'ma Zh.
  Eksp. Teor. Fiz. {\bf 66}, 811 (1997)].

\bibitem{Ustinov98a}
A. Ustinov and S. Sakai, Appl. Phys. Lett. {\bf 73},  686  (1998).

\bibitem{Volkov98}
A. Volkov and V. Glen, J. of Phys.: Cond. Matt. {\bf 10},  L563  (1998).

\bibitem{Machida00}
M. Machida, T. Koyama, A. Tanaka, and M. Tachiki, PHYSICA C {\bf 330},  85
  (2000).

\bibitem{Koshelev00}
A.~E. Koshelev and I.~S. Aranson, Phys. Rev. Lett {\bf 85},  3938  (2000).

\bibitem{Kleiner00}
R. Kleiner, T. Gaber, and G. Hechtfischer, Phys. Rev. B {\bf 62},  4086
  (2000); R. Kleiner, T. Gaber, and G. Hechtfischer, unpublished.

\bibitem{Bulaevskii96a}
L. Bulaevskii {\it et~al.}, Phys. Rev. B {\bf 53},  14601  (1996).

\bibitem{Bulaevskii96b}
L. Bulaevskii {\it et~al.}, Phys. Rev. B {\bf 54},  7521  (1996).

\bibitem{Koyama96}
T. Koyama and M. Tachiki, Phys. Rev. B {\bf 54},  16183  (1996).

\bibitem{Artemenko97}
S. Artemenko and A. Kobelkov, Phys. Rev. Lett. {\bf 78},  3551  (1997).

\bibitem{Ryndyk98}
D. Ryndyk, Phys. Rev. Lett. {\bf 80},  3376  (1998).

\bibitem{Ryndyk99}
D. Ryndyk, JETP {\bf 89},  975  (1999), [Zh. Eksp. Teor. Fiz. {\bf 116}, 1798
  (1999)].

\bibitem{Melnikov96}
A. Mel'nikov, Phys. Rev. Lett. {\bf 77},  2786-2789  (1996).

\end{thebibliography}
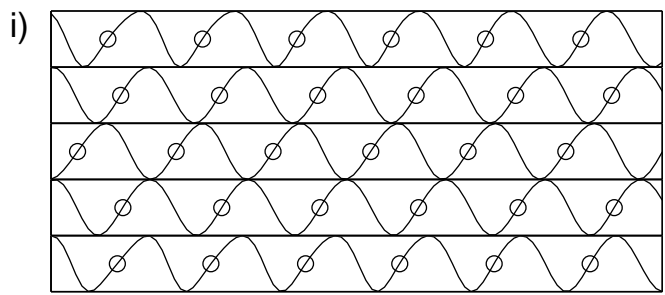
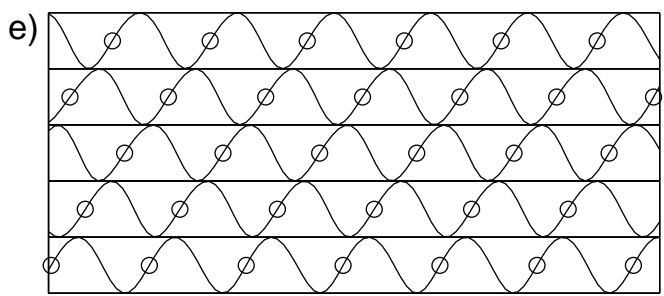
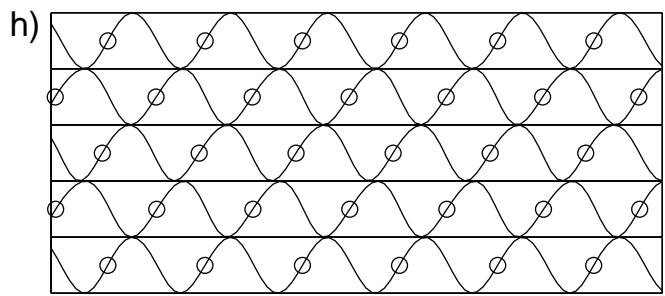
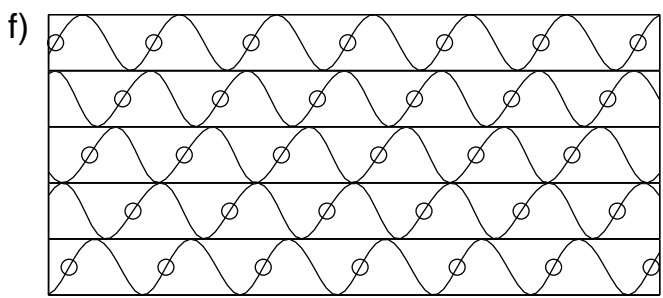
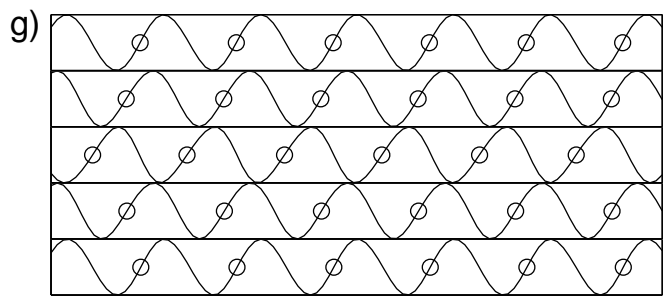
\end{document}